\documentclass[runningheads]{svmult}

\usepackage{makeidx}   
\usepackage{graphicx}  
\usepackage{subeqnar}  
\usepackage{physprbb}  
\makeindex             

\def\gsim{\mathrel{\raise0.35ex\hbox{$\scriptstyle >$}\kern-0.6em 
\lower0.40ex\hbox{{$\scriptstyle \sim$}}}}
\def\lsim{\mathrel{\raise0.35ex\hbox{$\scriptstyle <$}\kern-0.6em 
\lower0.40ex\hbox{{$\scriptstyle \sim$}}}}

\begin{document}
\title*{Galaxy Evolution in Three dimensions :\\ Time, Space and Mass}
\toctitle{Galaxy Evolution in Three dimensions : Time, Space and Mass}
%
%
\titlerunning{Galaxy Evolution in Three dimensions}
%
\author{Tadayuki Kodama\inst{1}
\and Richard Bower\inst{2}
\and Philip Best\inst{3}
\and Patrick Hall\inst{4}
\and Toru Yamada\inst{1}
\and Masayuki Tanaka\inst{5}}
\authorrunning{T. Kodama et al.}
%
%
\institute{National Astronomical Observatory of Japan, 2--21--1 Osawa, Mitaka, Tokyo, 181--8588, Japan
\and
Department of Physics, University of Durham, South Road, Durham DH1 3LE, UK
\and
Institute for Astronomy, Royal Observatory Edinburgh, Blackford Hill, Edinburgh EH9 3HJ
\and
Department of Astrophysical Sciences, Princeton University, Peyton Hall, Ivy Lane, Princeton, NJ 08544--1001, USA
\and
Department of Astronomy, School of Science, University of Tokyo,
Bunkyo-ku, Tokyo 113--0033, Japan}

\maketitle              

\begin{abstract}

There are three major axes to describe the evolution of galaxies,
namely, time (redshift), space (environment) and mass (stellar mass).
In this article, one topic each will be presented along these axes.
(1) Based on the Subaru wide-field (30$'$) optical imaging of two
distant clusters ($z$=0.55 and 0.83), we show the large scale structures
in galaxy distributions on a scale greater than 10~Mpc, which serve as
an evidence for hierarchical growth of rich clusters of galaxies
through assembly of surrounding groups.
(2) Based on the deep NIR imaging of high-$z$ clusters at $z\sim$1 and 1.5,
we show massive galaxies in clusters have assembled most of their mass
by $z=1.5$, which is earlier than the hierarchical model predictions.
(3) Based on the Subaru deep and wide optical imaging of
Subaru/XMM-Newton Deep Survey Field,
we show a deficit of red and faint galaxies
and a lack of blue massive galaxies
in the high density regions at $z\sim1$,
which suggest down-sizing in galaxy formation as apparently opposed to the
CDM-based bottom-up scenario.
\end{abstract}


\section{{\it Space}: Panoramic Imaging and
Spectroscopy of Cluster Evolution with Subaru (PISCES)}

\begin{figure}[t]
\begin{center}
\includegraphics[width=.56\textwidth]{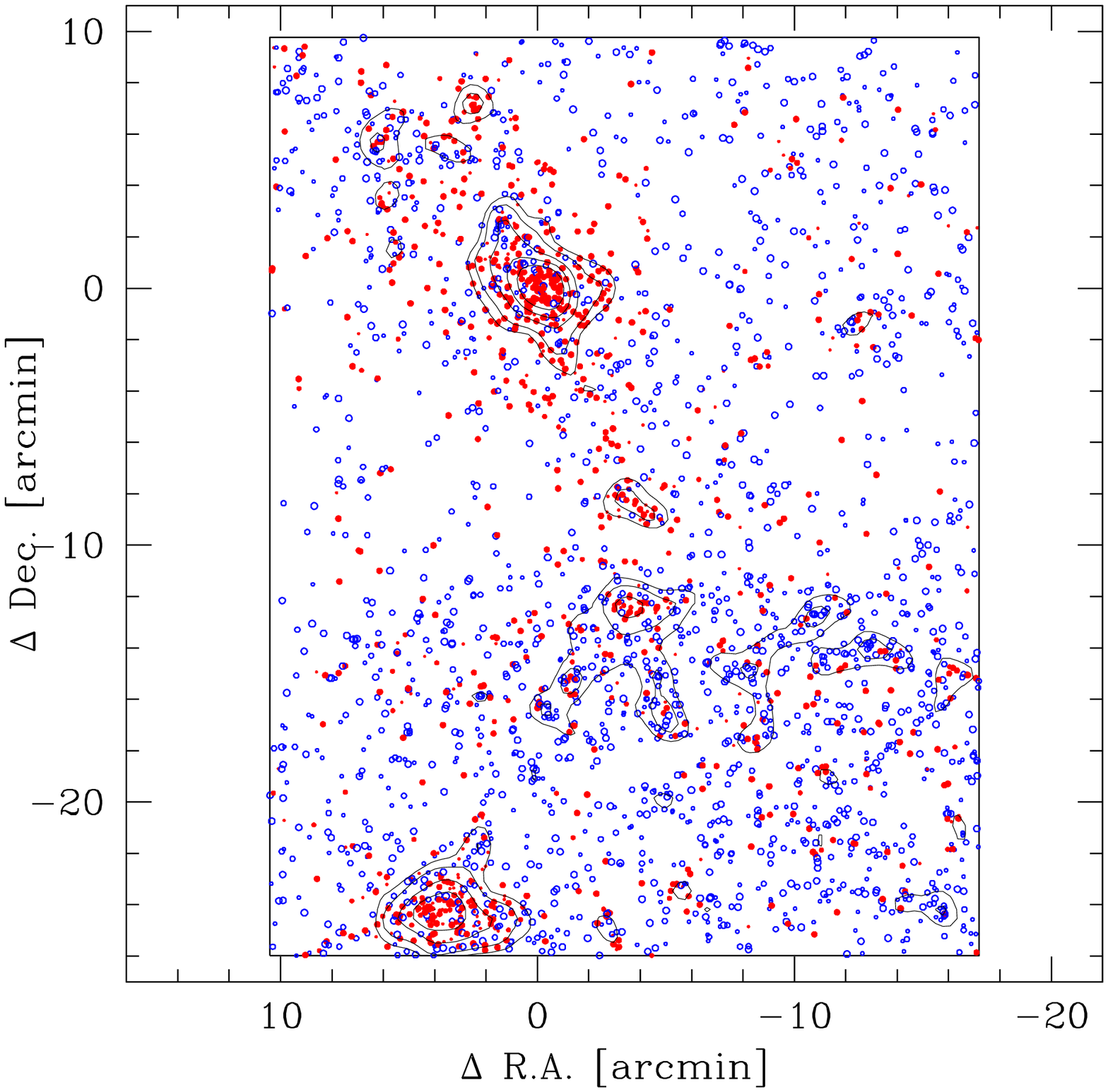}
\includegraphics[width=.56\textwidth]{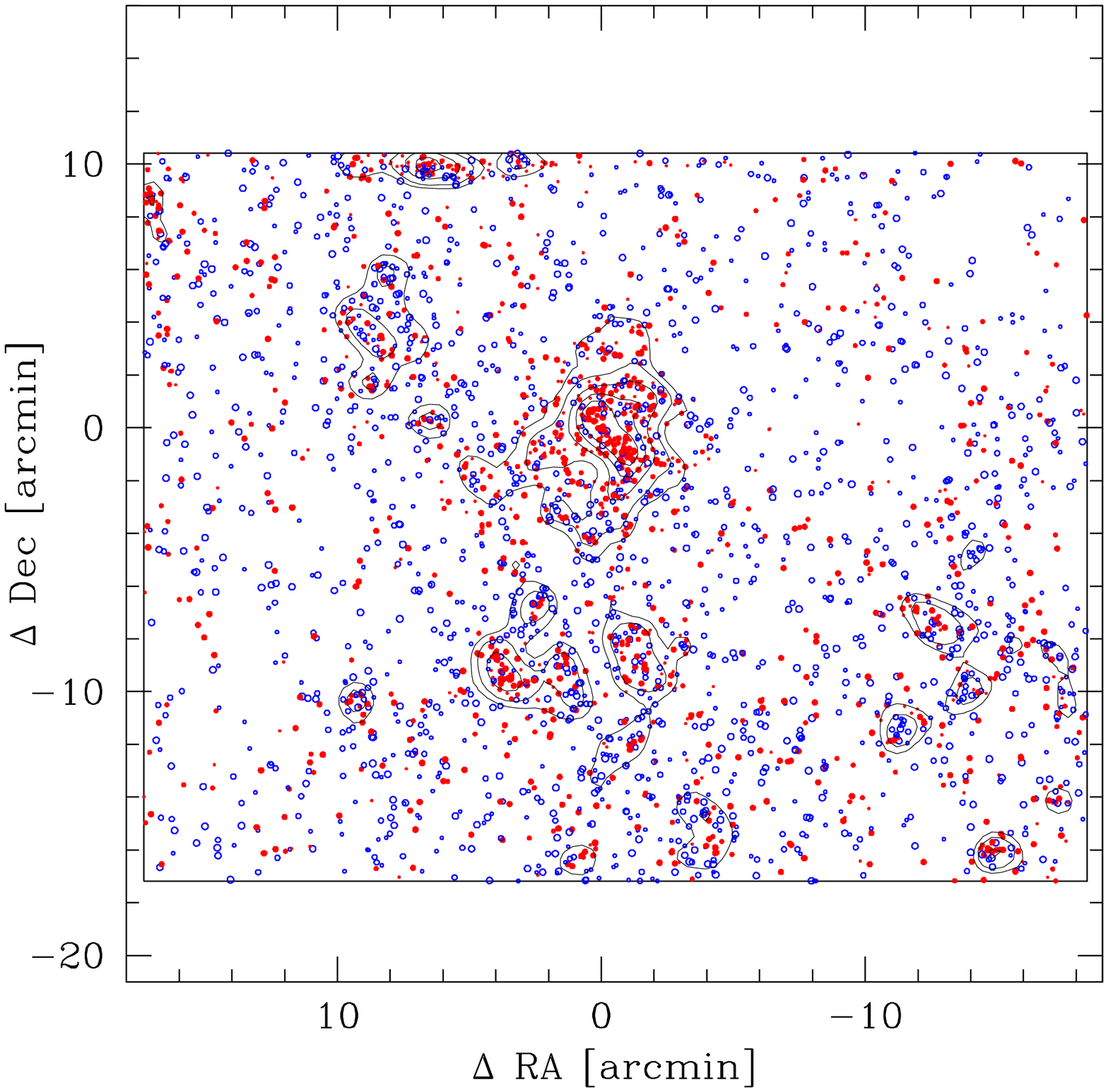}
\end{center}
\caption[]{The panoramic maps of CL0016+16 cluster
($z=0.55$) and RXJ0152.7--1357 ($z$=0.83) are presented on the top and the
bottom panels (10~arcminutes corresponds to 3.8~Mpc and 4.6~Mpc,
respectively).
Using photometric redshift technique based on multi-colour data
($BVRi'z'$ and $VRi'z'$, respectively), only photometric member
candidates are plotted selected with redshift cuts of 0.47$\le$$z$$\le$0.6
and 0.75$\le$$z$$\le$0.9, respectively.
Contours show local 2-D number density of galaxies at
1.5, 2, 3, 4, 5 $\sigma$ above the mean density.
Coordinates are shown relative to the centre of the main cluster.
Large scale filamentary structures ($>$10Mpc) are seen in both clusters.
}
\label{fig:rxj0153}
\end{figure}

First of all, along the space axis, we present panoramic imaging of
two distant clusters taken with Subaru wide-field optical camera
Suprime-Cam which provides 34$'$$\times$27$'$ field of view corresponding to
15$\times$12~Mpc$^2$ at $z=0.8$.
This project called PISCES (Panoramic Imaging and Spectroscopy of
Cluster Evolution with Subaru) has started since 2003, and
we aim to target 10-15 X-ray selected distant clusters in total at
$0.4\lsim z\lsim 1.3$, in good coordination with {\it ACS/HST},
{\it XMM}, and {\it Chandra} observations.
The goals of this programme is to map out the large scale structures
around clusters to trace the cluster assembly history and then to look into
the galaxy properties as a function of environment along the structures
to directly identify the environmental effects acting on galaxies
during their assembly to higher density regions.
This unique project is currently underway and just
a preliminary result on the large scale structures is shown in Fig.~1.
These two rich clusters at $z$=0.55 and 0.83 are imaged in multi optical
bands with Suprime-Cam, and photometric redshifts~\cite{k99}
have been applied to efficiently remove
foreground/background contaminations and to isolate
the cluster member candidates (cf., \cite{k01}).
Many substructures are now clearly seen around the main body of the
clusters which tend to be aligned in filamentary structures extending to
$>$10~Mpc scale across.
Although these structures should be confirmed spectroscopically later on,
these already provide good evidence for cluster assembly in the
hierarchical Universe.

\cite{k01} have presented the environmental dependence of galaxy
colours along the filamentary structures around the A851 cluster
($z$=0.41), and have shown that the galaxy colour changes rather sharply
at relatively low density regions like galaxy groups along the filaments
well outside of the cluster core.  Together with the similar findings
in the local Universe~\cite{lewis02}\cite{gomez03}.
the environmental effects that truncate star formation are not
the cluster specific phenomena such as ram-pressure stripping~\cite{abadi99}
but are found to be much wider spread into low density regions.
It is important to extend this analysis to higher redshifts as the
galaxy environment is expected to change dramatically during the course
of vigorous assembly, which is probably related to the appearence of
morphology-density relation~\cite{d80}.

\section{{\it Time:} Stellar mass assembly of massive galaxies in
high-$z$ clusters}

\begin{figure}[t]
\begin{center}
\includegraphics[width=.62\textwidth]{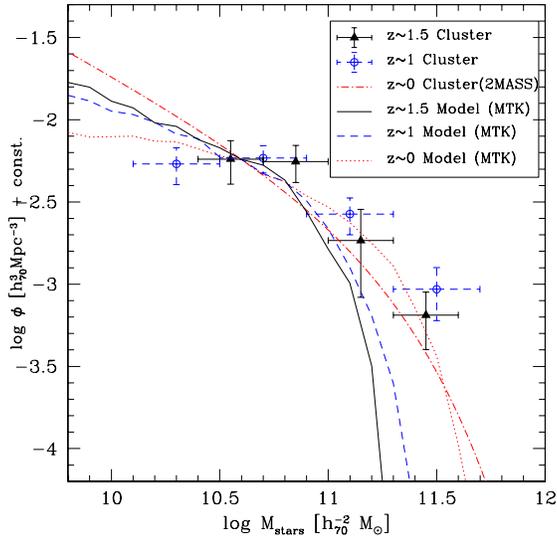}
\end{center}
\caption[]{Stellar mass functions of galaxies in clusters as a fuction of
redshift.
The Kennicutt's initial mass function~\cite{k83} is used to scale the
stellar mass.
The open diamonds and the filled triangles show
the stellar mass functions for $z\sim1$ and $z\sim1.5$ clusters,
respectively, which are compared to the local counterpart from the
2MASS survey (dot-dashed curve).
All the curves and the data points are normalised at
5$\times$10$^{10}$M$_{\odot}$ so as to have the same amplitude.
The theoretical predictions from a semi-analytic model~\cite{n02}
are presented for comparison, which are made for galaxies in the haloes
whose circular velosities are greater than 1000~km/s at each epoch. 
The mass assembly of massive galaxies in the real Universe is much
faster than the hierarchical model prediction.
}
\label{fig:mf}
\end{figure}

Secondly, along the time axis, we present the stellar mass
functions of galaxies in high-$z$ clusters constructed from deep
near-infrared imaging ($J$ and $K_s$) (Fig.~2).
The stellar mass function of galaixes derived from the $K$-band
observations is a good tracer of mass assembly history of galaxies
(eg., \cite{kc98}\cite{baugh02}).
We have combined two $z\sim1$ clusters (3C336 and Q1335+28)~\cite{kb03},
and five $z\sim1.5$ clusters (Q0835+580,
Q1126+101, Q1258+404, Q0139--273,
and Q2025--155)~\cite{hall98}\cite{hall01}\cite{best03}
to increase statistics. We have subtracted the control
field counts taken from the literature~\cite{s99}\cite{s01}\cite{best03}.
Applying the same technique described in \cite{kb03},
we construct the field-subtracted stellar mass functions
of galaxies in high-$z$ clusters, primarily using $K_s$-band flux and
also using $J-K_s$ colour as a measure of the M/L ratio.
As shown, little evolution is observed since $z=1.5$ to the present-day
(2MASS clusters~\cite{balogh01}),
indicating that the mass assembly on galaxy scale is largely completed
by $z\sim1.5$ in the cluster environment.
This epoch of mass assembly of massive galaxies is earlier than the
prediction of the hierarchical models as shown for comparison~\cite{n02}.

It is also interesting to trace back the stellar mass to even higher redshift
to firstly identify the epoch of assembly of massive galaxies where
they start to break down into pieces.
Recent deep NIR observations in fact start to enter such formation epoch
(eg., \cite{d03}\cite{r03}\cite{daddi03}),
and this aspect will be further extended by on-going and future space
missions such as {\it SIRTF} and {\it Astro-F}.

\section{{\it Mass:} Down-sizing in galaxy formation seen at $z\sim1$}

\begin{figure}[t]
\begin{center}
\includegraphics[width=.62\textwidth]{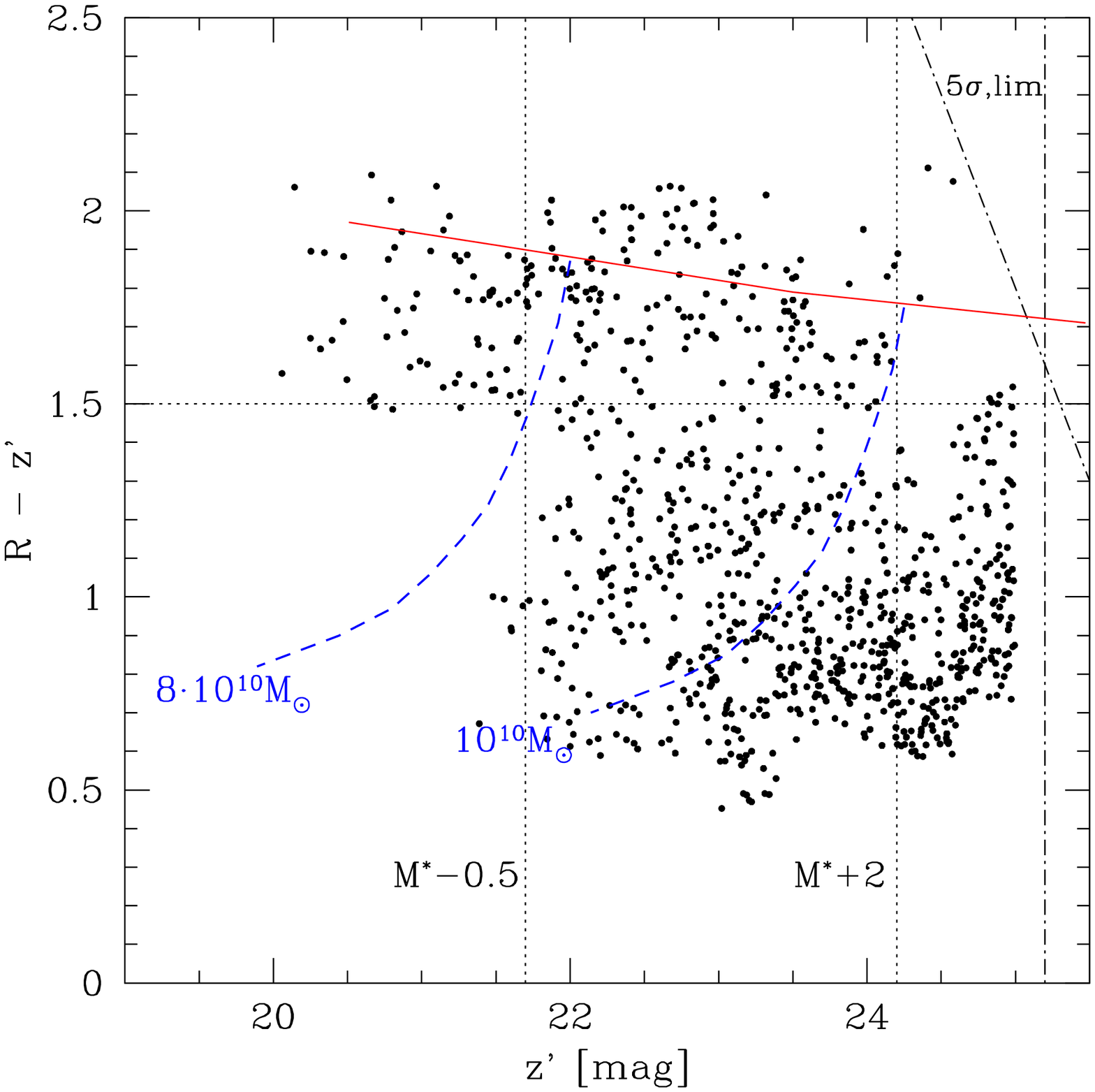}
\end{center}
\caption[]{Field-corrected colour-magnitude diagram for the $z\sim1$
galaxies in high density regions~\cite{ky03}.
The solid line show the expected location of colour-magnitude sequence
at $z\sim1$ assuming a passive evolution with $z_{\rm form}$=5~\cite{k98}.
The deficit of both blue galaxies at the
bright/massive end and the deficit of red galaxies at the faint/less-massive
end are both clearly identified.
}
\label{fig:z1cm}
\end{figure}

\begin{figure}[t]
\begin{center}
\includegraphics[width=.62\textwidth]{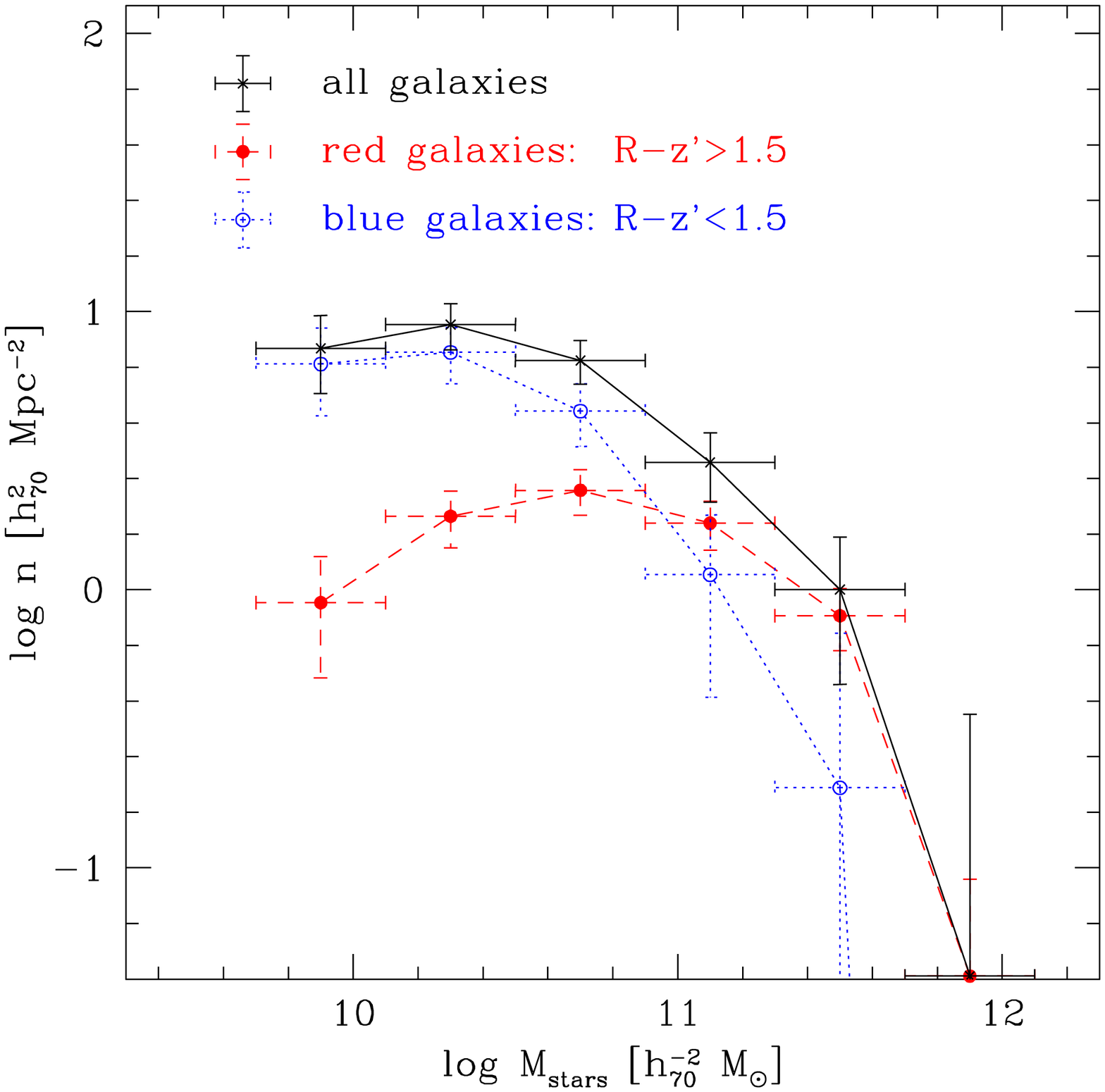}
\end{center}
\caption[]{Colour-dependent stellar mass functions of the $z\sim1$
galaxies in high density regions~\cite{ky03}.
The stellar masses are scaled using the Kennicutt's IMF~\cite{k83}.
}
\label{fig:mf}
\end{figure}

Finally, along the mass axis, we present how the star formation history
depends on the mass of the galaxies at $z\sim1$ based on the unique
Suprime-Cam imaging data ($BRi'z'$) on the Subaru/XMM-Newton Deep Survey
(SXDS) Field.
Our data are both sufficiently deep ($z'_{AB}$=25, 6-10$\sigma$) and
wide (1.2 deg$^2$), which enable us for the first time to investigate
the photometric properties of statistical sample of galaxies at $z\sim1$
down to $\sim$$M^*$+3 with respect to the passive evolution.
We first identify five $z\sim1$ high density regions by applying colour cuts
at 1.7$<$$R$$-$$z'$$<$2.0 and 0.8$<$$i'$$-$$z'$$<$1.1, which correspond
to the colours of passively evolving galaxies at $z\sim1$~\cite{k98}.
We then combine these five regions 
(amounting to 141 arcmin$^2$ in total) and subtract off the low density
regions at $z\sim1$ (scaled to the same area) on the colour-magnitude
diagram to isolate only $z\sim1$ galaxies in a statistical sense, since
the both regions are expected to have the same amount of foreground/background
contaminations.

Thus constructed field-corrected colour-magnitude diagram for $z\sim1$
galaxies in high density regions is shown in Fig.~3.
Most striking feature of this diagram is that there are two distinct
populations, `bright+red' and `faint+blue'.
More precisely, we show a deficit of red and faint galaxies
below $M^*$+2 or 10$^{10}$M$_{\odot}$ in stellar mass
and a lack of blue massive galaxies
beyond $M^*$$-$0.5 or 8$\times$10$^{10}$M$_{\odot}$ in stellar mass.
This bimodality in colour-magnitude distributions can be also seen in
the colour-dependent stellar mass function shown in Fig.~4.
Red galaxies tend to be bright and show a bell-like shape,
while the blue galaxies tend to be faint and show a much steeper
faint-end slope.
These results indicate the down-sizing in galaxy formation~\cite{cowie96},
where star formation and mass assembly in massive galaxies take place early
in the Universe,
while the star formation seems to be progressively shifted towards less
massive systems as the Universe ages.
Next obvious important step is to directly detect the brightening
of such break luminosity or mass with increasing redshift~\cite{kb01}.

\vspace{0.3cm}

We have assumed the cosmological parameters of
($H_0$, $\Omega_m$, $\Omega_{\Lambda}$)=(70, 0.3, 0.7), throughout this
paper.

\vspace{0.3cm}

%
%
We thank Drs. Ian Smail and Eric Bell for useful discussion.
We also acknowledge to the SXDS team (Kaz Sekiguchi et al.).
This work was financially supported in part by a Grant-in-Aid for the
Scientific Research (No.\, 15740126) by the Japanese Ministry of Education,
Culture, Sports and Science.

%

\end{document}